\documentstyle[12pt]{article}
\begin{document}
\noindent
\begin{center}
{\Large {\bf Solar System Constraints on\\ a Cosmologically Viable
f(R) Theory\\}} \vspace{2cm}

 ${\bf Yousef~
Bisabr}$\footnote{e-mail:~y-bisabr@srttu.edu.}\\
\vspace{.5cm} {\small{Department of Physics, Shahid Rajaee Teacher
Training University,
Lavizan, Tehran 16788, Iran.}}\\
\end{center}
\vspace{1cm}
\begin{abstract}
Recently, a model $f(R)$ theory is proposed \cite{recent} which is
cosmologically viable and distinguishable from $\Lambda$CDM.  We
use chameleon mechanism to investigate viability of the model in
terms of Solar System experiments.

\end{abstract}
 \vspace{3cm}
\section{Introduction}
There are strong observational evidences that the expansion of the
universe is accelerating. These observations are based on type Ia
supernova \cite{super}, cosmic microwave background radiation
\cite{cmbr}, large scale structure formation \cite{ls}, weak
lensing \cite{wl}, etc. The standard explanation invokes an
unknown component, usually referred to as dark energy. It
contributes to energy density of the universe with
$\Omega_{d}=0.7$ where $\Omega_{d}$ is the corresponding density
parameter, see e.g., \cite{ca} and references therein. The
simplest dark energy scenario which seems to be both natural and
consistent with observations is the $\Lambda$CDM model in which
dark energy is identified as a cosmological constant \cite{ca}
\cite{wei} \cite{cc}. However, in order to avoid theoretical
problems \cite{wei}, other scenarios have been investigated. Among
different scenarios there are modified gravity models \cite{modi},
in which one modifies the laws of gravity whereby a late time
acceleration is produced without recourse to a dark energy
component.  One family of these modified gravity models is
obtained by replacing the Ricci scalar $R$ in the usual
Einstein-Hilbert Lagrangian density for some function $f(R)$.
These models are cosmologically acceptable if they meet some
certain conditions.  The most important ones is that they should
follow a usual matter-dominated era preceding a late-time
accelerated stage.  Several models are proposed that admit
cosmological solutions with accelerated expansion of the universe
at late times \cite{acc}.  However, among all cosmologically
viable $f(R)$ theories there is still an important issue to be
pursued, they must be probed at Solar System scale. In fact,
changing gravity Lagrangian has consequences not only in
cosmological scales but also in galactic and Solar System ones so
that it seems to be necessary to investigate the low energy limit
of such $f(R)$ theories. \\ Early works on the weak field limit of
$f(R)$ theories led to negative results.  Using the equivalence of
$f(R)$ and scalar-tensor theories \cite{soko} \cite{maeda}
\cite{wands}, it is originally suggested that all $f(R)$ theories
should be ruled out \cite{chiba} since they violate the weak field
constraints coming from Solar System experiments. This claim was
based on the fact that $f(R)$ theories (in the metric formalism)
are equivalent to Brans-Dicke theory with $\omega=0$ while
observations set the constraint $\omega>40000$ \cite{will}.  In
this case the post-Newtonian parameter satisfies
$\gamma=\frac{1}{2}$ instead of being equal to unity as required
by observations.  Later, it was noted by many authors that for
scalar fields with sufficiently large mass it is possible to drive
$\gamma$ close to unity even for null Brans-Dicke parameter.  In
this case the scalar field becomes short-ranged and has no effect
at Solar System scales. Recently, it is shown that there exists an
important possibility that the effective mass of the scalar field
be scale dependent \cite{k}. In this chameleon mechanism, the
scalar field may acquire a large effective mass in Solar System
scale so that it hides local experiments while at cosmological
scales it is
effectively light and may provide an appropriate cosmological behavior.\\
In the present work we intend to use this criterion to investigate
constraints set by local experiments on a cosmologically viable
model recently proposed by Miranda et al \cite{recent}.  It is a
two parameter $f(R)$ model which is introduced in the form
\begin{equation}
f(R)=R-\alpha R_{1} \ln({1+\frac{R}{R_{1}})}
\label{r1}\end{equation} where $\alpha$ and $R_{1}$ are positive
parameters.  This model is reduced to general relativity for
$\alpha=0$.  A cosmologically viable $f(R)$ model must start with
a radiation-dominated era and have a saddle point matter-dominated
phase followed by an accelerated epoch as a final attractor.  This
is formally stated by introducing the parameters
$m=R\frac{d^2f/dR^2}{dR^2}$ and $r=-R\frac{df/dR}{f}$ \cite{amen}.
The cosmological dynamics of $f(R)$ models can then be understood
by considering $m(r)$ curves in the $(r, m)$ plane. Using this
criteria, the authors of \cite{recent} showed that the model
(\ref{r1}) satisfies all the cosmological requirements for
$\alpha>1$ and regardless of $R_{1}$ \footnote{In the model
proposed in \cite{recent}, it is stated that the relation
(\ref{r1}) is a special case of a general parametrization unifying
the models of \cite{saw} and \cite{star}.  In the latter models
there is a parameter with the same dimension of $R_{1}$ which is
taken to be of the same order of the presently observed
cosmological constant. The relation between this parameter and
$R_{1}$ and also the relevance of  $R_{1}$ with the time of the
beginning of the acceleration phase are important issues that are
not explicitly addressed in \cite{recent}.} Here we will focus on
viability of the model in terms of local gravity experiments.  We
will show that these local experiments
rule out the model as an explanation for the current accelerated expansion of the universe.\\
\section{Chameleon Mechanism}
In this section we offer a brief review of the chameleon
mechanism. We consider the following action\footnote{We use the
unit $(8\pi G)^{-1}=1.$}
\begin{equation}
S=\frac{1}{2} \int d^{4}x \sqrt{-g}~ f(R) + S_{m}(g_{\mu\nu},
\psi)\label{a}\end{equation} where $g$ is the determinant of
$g_{\mu\nu}$, $f(R)$ is an unknown function of the scalar
curvature $R$ and $S_{m}$ is the matter action depending on the
metric $g_{\mu\nu}$ and some matter field $\psi$. We may use a new
set of variables
\begin{equation}
\bar{g}_{\mu\nu} =p~ g_{\mu\nu} \label{c}\end{equation}
\label{a1}\begin{equation} \phi = \frac{1}{2\beta} \ln p
\label{dd}\end{equation}
 where
$p\equiv\frac{df}{dR}=f^{'}(R)$ and $\beta=\sqrt{\frac{1}{6}}$.
This is indeed a conformal transformation which transforms the
above action in the Jordan frame to the Einstein frame \cite{soko}
\cite{maeda} \cite{wands}
\begin{equation}
S=\frac{1}{2} \int d^{4}x \sqrt{-g}~\{ \bar{R}-\bar{g}^{\mu\nu}
\partial_{\mu} \phi~ \partial_{\nu} \phi -2V(\phi)\} + S_{m}(\bar{g}_{\mu\nu}
e^{2\beta \phi}, \psi) \label{s}\end{equation} In the Einstein
frame, $\phi$ is a scalar field with a self-interacting potential
which is given by
\begin{equation}
V(\phi)=\frac{1}{2}e^{-2\beta \phi} \{r[p(\phi)]-e^{-2\beta \phi}
f(r[p(\phi)])\} \label{v}\end{equation} where $r(p)$ is a solution
of the equation $f^{'}[r(p)]-p=0$ \cite{soko}.  Note that
conformal transformation induces the coupling of the scalar field
$\phi$ with the matter sector. The strength of this coupling
$\beta$, is fixed to be $\sqrt{\frac{1}{6}}$ and is the same for
all types of matter fields.  In the case of such a strong matter
coupling, the role of the potential of the scalar field is
important for consistency with local gravity experiments. When the
potential satisfies certain conditions it is possible to attribute
an effective mass to the scalar field which has a strong
dependence on ambient density of matter.  A theory in which such a
dependence is realized is said to be a chameleon theory \cite{k}.
In such a theory the scalar field $\phi$ can be heavy enough in
the environment of the laboratory tests so that the local gravity
constraints suppressed even if $\beta$ is of the order of unity.
Meanwhile, it can be light enough in the low-density cosmological
environment to be
considered as a candidate for dark energy.\\
Variation of the action (\ref{a}) with respect to
$\bar{g}_{\mu\nu}$ and $\phi$, gives the field equations
\begin{equation}
\bar{G}_{\mu\nu}=\partial_{\mu} \phi~\partial_{\nu} \phi
-\frac{1}{2}\bar{g}_{\mu\nu} \partial_{\gamma}
\phi~\partial^{\gamma} \phi-V(\phi)
\bar{g}_{\mu\nu}+\bar{T}_{\mu\nu} \label{g}\end{equation}
\begin{equation}
\bar{\Box} \phi -\frac{dV}{d\phi}=-\beta \bar{T}
\label{b}\end{equation} where
\begin{equation}
\bar{T}_{\mu\nu}=\frac{-2}{\sqrt{-g}}\frac{\delta S_{m}}{\delta
\bar{g}^{\mu\nu}}\label{t}\end{equation} and
$\bar{T}=\bar{g}^{\mu\nu}\bar{T}_{\mu\nu}$.  Covariant
differentiation of (\ref{g}) and the Bianchi identities give
\begin{equation}
\bar{\nabla}^{\mu} \bar{T}_{\mu\nu}=\beta~\bar{T}~\partial_{\nu}
\phi \label{b1}\end{equation} which implies that the matter field
is not generally conserved and feels a new force due to gradient
of the scalar field.  Let us consider $\bar{T}_{\mu\nu}$ as the
stress-tensor of dust with energy density $\bar{\rho}$ in the
Einstein frame. In a static and spherically symmetric spacetime
the equation (\ref{b}) gives
\begin{equation}
\frac{d^2 \phi}{d\bar{r}^2}+\frac{2}{\bar{r}}
\frac{d\phi}{d\bar{r}}=\frac{dV_{eff}(\phi)}{d\phi}
\label{d}\end{equation} where $\bar{r}$ is distance from center of
the symmetry in the Einstein frame and
\begin{equation}
V_{eff}(\phi)=V(\phi)-\frac{1}{4}\rho e^{-4\beta \phi}
\label{d1}\end{equation} Here we have used the relation
$\bar{\rho}=e^{-4\beta \phi} \rho$ that relates the energy
densities in the Jordan and the Einstein frames.  We consider a
spherically symmetric body with a radius $\bar{r}_{c}$ and a
constant energy density $\bar{\rho}_{in}$ ($\bar{r}<
\bar{r}_{c}$). We also assume that the energy density outside the
body ($\bar{r}> \bar{r}_{c}$) is given by $\bar{\rho}_{out}$. We
will denote by $\varphi_{in}$ and $\varphi_{out}$ the field values
at two minima of the effective potential $V_{eff}(\phi)$ inside
and outside the object, respectively. They must clearly satisfy
$V^{'}_{eff}(\varphi_{in})=0$ and $V^{'}_{eff}(\varphi_{out})=0$
where prime indicates differentiation of $V_{eff}(\phi)$ with
respect to the argument.  As usual, masses of small fluctuations
about these minima are given by
$m_{in}=[V^{''}_{eff}(\varphi_{in})]^{\frac{1}{2}}$ and
$m_{out}=[V^{''}_{eff}(\varphi_{out})]^{\frac{1}{2}}$ which depend
on ambient matter density. A region with large mass density
corresponds to a heavy mass field while regions with low mass
density corresponds to a field with lighter mass. In this way it
is possible for the mass field to take sufficiently large values
near massive objects in the Solar System scale and to hide the
local tests. For a spherically symmetric body there is a
thin-shell condition
\begin{equation}
\frac{\Delta
\bar{r}_{c}}{\bar{r}_{c}}=\frac{\varphi_{out}-\varphi_{in}}{6\beta
\Phi_{c}}\ll 1 \label{l1}\end{equation} where $\Phi_{c}=M_{c}/8\pi
\bar{r}_{c}$ is the Newtonian potential at $\bar{r}=\bar{r}_{c}$
with $M_{c}$ being the mass of the body.  In this case, equation
(\ref{d}) with some appropriate boundary conditions gives the
field profile outside the object \cite{k}
\begin{equation}
\phi(\bar{r})=-\frac{\beta}{4\pi} \frac{3\Delta
\bar{r}_{c}}{\bar{r}_{c}}\frac{M_{c} e^{-m_{out}
(\bar{r}-\bar{r}_{c})}}{\bar{r}}+\varphi_{out}
\label{l2}\end{equation}
~~~~~~~~~~~~~~~~~~~~~~~~~~~~~~~~~~~~~~~~~~~~~~~~~~~~~~~~~~~~~~~~~~~~~~~~~~~~~~~~~~~~~~~~
\section{The Model}
The function $f(R)$ in the Jordan frame is closely related to the
potential function of the scalar degree of freedom of the theory
in the Einstein frame.  Any functional form for the potential
function corresponds to a particular class of $f(R)$ theories. To
find a viable function $f(R)$ passing Solar System tests one can
equivalently work with its corresponding potential function in the
Einstein frame and put constraints on the relevant parameters via
chameleon mechanism. Taking this as our criterion, we write
potential function of the model (\ref{r1})
\begin{equation}
V(\phi)=\frac{1}{2} R_{1} e^{-4\beta \phi}\{\alpha
\ln({\frac{\alpha}{1-e^{2\beta \phi}}})-e^{2\beta
\phi}-(\alpha-1)\} \label{11}\end{equation}
 Assuming that $\phi<<1$, one
can find the solution of $V^{'}_{eff}(\phi)=0$ by substituting
(\ref{11}) into (\ref{d1})
\begin{equation}
\varphi=\frac{1}{2A}\{-C\pm \sqrt{C^2 -4AB}\}
\label{n}\end{equation} where
\begin{equation}
A=2\beta^2 R_{1}(1-2\alpha)
\end{equation}
\begin{equation}
B=-\frac{1}{2}\alpha R_{1}
\end{equation}
\begin{equation}
C=\beta [\rho+R_{1}(\alpha-1)-2\alpha R_{1}\ln{\alpha}]
\end{equation}
In the following we shall consider thin-shell condition together
with constraints set by equivalence principle and fifth force experiments.\\\\
$1.~ Thin-shell~~ condition$\\
In the chameleon mechanism, the chameleon field is trapped inside
large and massive bodies and its influence on the other bodies is
only due to a thin-shell near the surface of the body.  The
criterion for this thin-shell condition is given by (\ref{l1}). If
we combine (\ref{l1}) and (\ref{n}) we obtain
\begin{equation}
\frac{\Delta \bar{r}_{c}}{\bar{r}_{c}}=\frac{1}{12\Phi_{c}A}
\{(\rho_{in}-\rho_{out})\pm \sqrt{(\rho_{out}+a)^2 -b} \mp
\sqrt{(\rho_{in}+a)^2-b}~\} \label{h}\end{equation} where
$\rho_{in}$ and $\rho_{out}$ are energy densities inside and
outside of the body in the Jordan frame.  Here
$a=\frac{A}{\beta}-\rho$ and $b=4\alpha R_{1}^2(2\alpha-1)$.  In
weak field approximation, the spherically symmetric metric in the
Jordan frame is given by
\begin{equation}
ds^{2}=-[1-2X(r)]dt^{2}+[1+2Y(r)]dr^{2}+r^2 d\Omega^2
\end{equation}
where $X(r)$ and $Y(r)$ are some functions of $r$. There is a
relation between $r$ and $\bar{r}$ so that $\bar{r}=p^{1/2} r$. If
we consider this relation under the assumption $ m_{out}~r\ll 1 $,
namely that the Compton wavelength $m_{out}^{-1}$ is much larger
than Solar System scales, then we have $\bar{r} \approx r$. In
this case, the chameleon mechanism gives for the post-Newtonian
parameter $\gamma$ \cite{faulk}
\begin{equation}
\gamma=\frac{3-\frac{\Delta r_{c}}{r_{c}}}{3+\frac{\Delta
r_{c}}{r_{c}}} \simeq 1-\frac{2}{3}\frac{\Delta r_{c}}{r_{c}}
\label{gg}\end{equation}\\
We can now apply (\ref{h}) on the Earth and obtain the condition
that the Earth has a thin-shell. To do this, we assume that the
Earth is a solid sphere of radius $R_{e} =6.4\times 10^{8}~cm$ and
mean density $\rho_{e} \sim 10~gr/cm^{3}$. We also assume that the
Earth is surrounded by an atmosphere with homogenous density
$\rho_{a} \sim 10^{-3}~gr/cm^{3}$ and thickness $100 km$.  For
simplifying equation (\ref{h}), we proceed under the assumption
$\rho_{in}, \rho_{out}<< R_{1}\alpha$\footnote{It can be easily
checked that our main results do not change when $\rho_{in},
\rho_{out}>> R_{1}\alpha$.}. We will return to this issue later.
In this case, equation (\ref{h}) simplifies to
\begin{equation}
\frac{\Delta
R_{e}}{R_{e}}\approx\frac{1}{4\Phi_{e}}\frac{(\rho_{in}-\rho_{out})}{R_{1}(1-2\alpha)}
\label{hhh}\end{equation} where $\Phi_{e}=6.95\times 10^{-10}$
\cite{wein} is Newtonian potential on surface of the Earth.  The
tightest Solar System constraint on $\gamma$ comes from Cassini
tracking which gives $\mid \gamma -1 \mid < 2.3 \times 10^{-5}$
\cite{will}. This together with (\ref{gg}) and (\ref{hhh}) yields
\begin{equation} (1-2\alpha)> 10^{13}(\frac{\rho_{in}}{R_{1}})\label{n+1}\end{equation}
With $\rho_{in}=\rho_{e}=7\times 10^{-28}~cm^{-2}$, this is
equivalent to
$R_{1}(1-2\alpha)> 10^{-15}~cm^{-2}$.\\\\
2. $Equivalence ~ principle$\\
We now consider constraints coming from possible violation of weak
equivalence principle. We assume that the Earth, together with its
surrounding atmosphere, is an isolated body and neglect the effect
of the other compact objects such as the Sun, the Moon and the
other planets. Far away the Earth, matter density is modeled by a
homogeneous gas with energy density $\rho_{G}\sim 10^{-24}
gr/cm^{3}$.  To proceed further, we first consider the condition
that the atmosphere of the Earth satisfies the thin-shell
condition \cite{k}. If the atmosphere has a thin-shell the
thickness of the shell ($\Delta R_{a}$) must be clearly smaller
than that of the atmosphere itself, namely $\Delta R_{a}< R_{a}$,
where $R_{a}$ is the outer radius of the atmosphere. If we take
thickness of the shell equal to that of the atmosphere itself
$\Delta R_{a}\sim 10^{2}~km$ we obtain $\frac{\Delta
R_{a}}{R_{a}}<1.5\times 10^{-2}$. It is then possible to relate
$\frac{\Delta R_{e}}{R_{e}}=\frac{\varphi_{a}-\varphi_{e}}{6\beta
\Phi_{e}}$ and $\frac{\Delta
R_{a}}{R_{a}}=\frac{\varphi_{G}-\varphi_{a}}{6\beta \Phi_{a}}$
where $\varphi_{e}$, $\varphi_{a}$ and $\varphi_{G}$ are the field
values at the local minimum of the effective potential in the
regions $r<R_{e}$ , $R_{a}>r>R_{e}$ and $r>R_{a}$ respectively.
Using the fact that newtonian potential inside a spherically
symmetric object with mass density $\rho$ is $\Phi \propto\rho
R^{2}$, one can write $\Phi_{e}=10^{4}~\Phi_{a}$ where $\Phi_{e}$
and $\Phi_{a}$ are Newtonian potentials on the surface of the
Earth and the atmosphere, respectively. This gives $\Delta R_{e}/
R_{e} \approx 10^{-4}~\Delta R_{a}/ R_{a}$. With these results,
the condition for the atmosphere to have a thin-shell is
\begin{equation}
\frac{\Delta R_{e}}{R_{e}} < 1.5\times 10^{-6}
\label{R}\end{equation}\\
The tests of equivalence principle measure the difference of
free-fall acceleration of the Moon and the Earth towards the Sun.
The constraint on the difference of the two acceleration is given
by \cite{will}
\begin{equation}
\frac{|a_{m}-a_{e}|}{a_{N}} < 10^{-13} \label{f}\end{equation}
where $a_{m}$ and $a_{e}$ are acceleration of the Moon and the
Earth respectively and $a_{N}$ is the Newtonian acceleration.  The
Sun and the Moon are all subject to the thin-shell condition
\cite{k} and the field profile outside the spheres are given by
(\ref{l2}) with replacement of corresponding quantities.  The
accelerations $a_{m}$ and $a_{e}$ are then given by \cite{k}
\begin{equation}
a_{e}\approx a_{N}\{1+18\beta^2 (\frac{\Delta R_{e}}{R_{e}})^2
\frac{\Phi_{e}}{\Phi_{s}}\}
\end{equation}
\begin{equation}
a_{m}\approx a_{N}\{1+18\beta^2 (\frac{\Delta R_{e}}{R_{e}})^2
\frac{\Phi_{e}^2}{\Phi_{s}\Phi_{m}}\}
\end{equation}
where $\Phi_{e}=6.95\times 10^{-10}$, $\Phi_{m}=3.14\times
10^{-11}$ and $\Phi_{s}=2.12\times 10^{-6}$ are Newtonian
potentials on the surfaces of the Earth, the Moon and the Sun,
respectively \cite{wein}. This gives a difference of free-fall
acceleration
\begin{equation}
\frac{|a_{m}-a_{e}|}{a_{N}}=(0.13)~ \beta^2~(\frac{\Delta
R_{e}}{R_{e}})^2
\end{equation}
Combining this with (\ref{f}) results in
\begin{equation}
\frac{\Delta R_{e}}{R_{e}} < 6.74\times 10^{-6}
\label{RR}\end{equation} which is of the same order of the
condition (\ref{R}) that the atmosphere has a thin-shell.  Taking
this as the constraint coming from violation of equivalence
principle and combining with (\ref{hhh}), we obtain
\begin{equation}
(1-2\alpha)>(10^{14})
(\frac{\rho_{in}}{R_{1}})\label{nn}\end{equation} which is not much different from (\ref{n+1}).\\\\
3. $Fifth~force$\\
The potential energy associated with a fifth force interaction is
parameterized by a Yukawa potential
\begin{equation}
U(r)=-\varepsilon\frac{m_{1}m_{2}}{8\pi}\frac{e^{-r/\lambda}}{r}
\label{y}\end{equation} where $m_{1}$ and $m_{2}$ are masses of
the two test bodies separating by distance r, $\varepsilon$ is
strength of the interaction and $\lambda$ is the range.  Thus
fifth force experiment constrains regions of ($\varepsilon,
\lambda$) parameter space. These experiments are usually carried
out in a vacuum chamber in which the range of the interaction
inside it is of the order of the size of the chamber \cite{k},
namely $\lambda\sim R_{vac}$. The tightest bound on the strength
of the interaction is $\varepsilon<10^{-3}$ \cite{h}.  Inside the
chamber we consider two identical bodies with uniform densities
$\rho_{c}$, radii $r_{c}$ and masses $m_{c}$.  If the two bodies
satisfy the thin-shell condition, their field profile outside the
bodies are given by
\begin{equation}
\phi(r)=-\frac{\beta}{4\pi} \frac{3\Delta r_{c}}{r_{c}}\frac{m_{c}
~e^{-r/R_{vac}}}{r}+\varphi_{vac} \label{22}\end{equation} Then
the corresponding potential energy of the interaction is
\begin{equation}
V(r)=-2\beta^2 (\frac{3\Delta
r_{c}}{r_{c}})^{2}~\frac{m_{c}^2}{8\pi}\frac{e^{-r/R_{vac}}}{r}
\label{yy}\end{equation} The bound on the strength of the
interaction translates into
\begin{equation}
2\beta^2 (\frac{3\Delta r_{c}}{r_{c}})^{2}<10^{-3}
\label{222}\end{equation} One can write for each of the test
bodies
\begin{equation}
\frac{\Delta
r_{c}}{r_{c}}\approx\frac{1}{4\Phi_{c}}\frac{(\rho_{c}-\rho_{vac})}{R_{1}(1-2\alpha)}
\label{hh}\end{equation} where $\rho_{vac}$ is energy density of
the vacuum inside the chamber.  In the experiment carried out in
\cite{h}, one used a typical test body with mass $m_{c}\approx 40
gr$ and radius $r_{c}\approx 1 cm$. These correspond to $\rho_{c}
\approx 9.5 gr/cm^{3}$ and $\Phi_{c} \sim 10^{-27}$. Moreover, the
pressure in the vacuum chamber was reported to be $3\times
10^{-8}$ Torr which is equivalent to $\rho_{vac} \approx 4.8\times
10^{-14}~gr/cm^{3}$. Substituting these into (\ref{hh}) and
combining the result with (\ref{222}) gives the bound
\begin{equation}
(1-2\alpha)>(10^{27})(\frac{\rho_{c}}{R_{1}})
\label{21}\end{equation} which is equivalent to $R_{1}(1-2\alpha)>10^{-1}~cm^{-2}$.\\
~~~~~~~~~~~~~~~~~~~~~~~~~~~~~~~~~~~~~~~~~~~~~~~~~~~~~~~~~~~~~~~~~~~~~~~~~~~~~~~~~~~~~~
\section{Discussion}
We have discussed viability of the $f(R)$ model proposed in
\cite{recent} in terms of local gravity constraints. We have used
the correspondence between a general $f(R)$ theory with scalar
field theories.  In general, in the scalar field representation of
a $f(R)$ theory there is a strong coupling of the scalar field
with the matter sector. We have considered the conditions that
this coupling is suppressed by chameleon mechanism.  We have found
that in order that the model (\ref{r1}) be consistent both with
fifth force and equivalence principle experiments, the two
parameters $\alpha$ and $R_{1}$ together should satisfy the
condition $R_{1}(1-2\alpha)>10^{-1}~cm^{-2}$. To have a bound on
the parameter $\alpha$, one should attribute a physical meaning to
the dimensional quantity $R_{1}$. Following the models proposed by
Hu and Sawicki \cite{saw} and Starobinski \cite{star}, if we take
it as the same order of the observed cosmological constant
$\Lambda_{obs}\sim
10^{-58}~cm^{-2}$ we obtain $|2\alpha-1|>10^{57}$.\\
It should be pointed out that this result is obtained under the
assumption $\rho_{in}, \rho_{out}<< R_{1}\alpha$. To understand
the relevance of this assumption, let us consider the case that
$\rho_{in}, \rho_{out}\sim R_{1}\alpha$. Taking energy density
inside the Earth as a typical energy density in the Solar System,
we obtain $R_{1}\alpha\sim 10^{-28}~cm^{-2}$. In the model
(\ref{r1}), if the coefficient $R_{1}\alpha$ is so small then it
would be hardly distinguishable from general relativity.  As the
point of view of local gravity experiments, a viable $f(R)$ model
should simultaneously satisfy Solar System bounds as well as
exhibit an appropriate deviation from general relativity.  This
requires that $R_{1}\alpha
>> 10^{-28}~cm^{-2}$, which confirms
the assumption that our results are based on.\\
The last point we wish to remark is that, as reported by the
authors of \cite{recent}, the condition that the universe pass
through a matter-dominated epoch and finally reach a late-time
accelerated phase is that $\alpha>1$ regardless of $R_{1}$.  This
seems not to be consistent with our results in the context of the
chameleon mechanism. Although our analysis do not place any
experimental bound on the parameter $R_{1}$, however for $R_{1}>0$
the relation (\ref{21}) implies that $\alpha \in (-\infty ~
\frac{1}{2}]$ which is out of the range reported in
\cite{recent}.\\\\
{\bf Acknowledgment}\\
The author would like to acknowledge the anonymous referee for
useful comments.

\end{document}